\documentclass[a4paper,12pt]{article}
\usepackage[T1]{fontenc}
\usepackage[utf8]{inputenc}
\usepackage{lmodern}
\usepackage{graphicx}
\usepackage{epstopdf} 
\usepackage{color}
\usepackage{bbold}
\usepackage{amsmath}
\title{Non-abelian field cohomology, its relation with spontaneous symmetry breaking and Morse's Theorem}

\author{ 
	V. E. R. Lemes$^{a}$\footnote{email: verlemes@gmail.com},
	 \\
	\small \em $^a$Instituto de F\'\i sica, Universidade do Estado do Rio de
	Janeiro,\\
	\small \em Rua S\~{a}o Francisco Xavier 524, Maracan\~{a}, Rio de Janeiro - RJ,
	20550-013, Brazil\\
	}

\begin{document}
	\maketitle
	\begin{abstract}	
		We show that, for an $SU(2)$ gauge field (the reasoning extends trivially to $SU(N)$), spontaneous symmetry breaking changes the field cohomology. This defines a new field with cohomological properties characteristic of matter fields. Consequently, the construction of a renormalizable unitary gauge fixing, following Morse's problem of functional extremization, leads to the Gribov condition being automatically solved on-shell. This result occurs because a specific combination of fields is cohomologically matter-like and therefore free of the Gribov problem.
	\end{abstract}

\section{Introduction}

The Gribov problem \cite{Gribov:1977wm} is a well-known obstruction to the unique gauge fixing of non-abelian gauge theories. It arises because the gauge fixing condition, such as the Landau gauge $\partial_\mu A_\mu = 0$, admits multiple solutions (Gribov copies) related by large gauge transformations. This ambiguity complicates the path integral quantization and challenges the standard perturbative framework.

One way to circumvent the Gribov problem is to restrict the gauge field configuration space to the Gribov region, where the Faddeev-Popov operator is positive. However, this restriction does not completely eliminate copies on the boundary, and the region itself is not convex. Alternative approaches, such as the Gribov-Zwanziger action, introduce non-local terms to account for the horizon condition. Nevertheless, these methods often involve explicit symmetry breaking or non-perturbative modifications whose renormalization properties are intricate.

In this work, we take a different perspective. We analyze the problem through the lens of algebraic renormalization and BRST cohomology \cite{Piguet:1995er}. It is known that spontaneous symmetry breaking changes the field content of the theory: certain combinations of gauge and scalar fields acquire cohomological properties characteristic of matter fields rather than gauge fields. We show that, in such a situation, a suitable gauge fixing constructed from a Morse functional automatically solves the Gribov problem on-shell.

The Morse problem, originally formulated for functionals on manifolds, concerns the existence and uniqueness of critical points. In gauge theories, the gauge fixing condition can be interpreted as the Euler-Lagrange equation of a functional of the gauge field (and possibly of scalar fields). When this functional is of Morse type, its critical points are isolated and non-degenerate. We argue that, in the broken phase, the on-shell Gribov operator acquires a positive mass term that lifts the zero modes responsible for the Gribov copies.

To make this argument precise, we study the cohomological structure of the theory before and after spontaneous symmetry breaking. The main technical tool is the filtration theorem \cite{Piguet:1995er}, which states that the cohomology of the full BRST operator $s$ is a subspace of the cohomology of a filtered operator $s_0$. This filtered operator captures the linearized symmetry transformations in the broken vacuum and reveals the emergence of matter-like field combinations that are invariant under $s_0$. These combinations can be lifted to full $s$-invariant observables.

We illustrate the mechanism with the electroweak sector $SU(2)\times U(1)$, spontaneously broken to $U(1)_{\text{em}}$. The extension to $SU(N)$ is straightforward. We construct the most general gauge fixing functional of ultraviolet dimension $2$ and ghost number zero, containing gauge fields, ghosts and scalar fields. We then show that, in the broken phase, the on-shell Gribov equation contains a mass term proportional to the symmetry breaking scale. For energies below that scale, the Gribov operator becomes positive definite, eliminating copies on-shell.

The paper is organized as follows. We first review the $SU(2)\times U(1)$ model with spontaneous symmetry breaking and compute the filtered BRST transformations. We then analyze the cohomology of the filtered operator, identify matter-like combinations, and introduce the Morse functional approach to derive the on-shell Gribov equation. The emergence of the mass term and its consequences for the Gribov problem are discussed. We conclude with a brief discussion of the implications for renormalizability and unitarity.

\section{Cohomology and spontaneous symmetry breaking: $SU(2)\times U(1)$ example}

To study cohomology in non-abelian gauge theory \cite{Piguet:1995er}, we consider a well-known example, $SU(2)\times U(1)$. We have two complex scalar fields in the fundamental representation. These fields transform under the BRST symmetry as:

\begin{eqnarray}
s\phi_{i}&=& -ig (T^{a})_{ij}c^{a}\phi_{j} - iec\phi_{i}\nonumber \\
s\phi^{\dagger}_{i}&=& ig \phi^{\dagger}_{j}c^{a}(T^{a})_{ji}+iec\phi^{\dagger}_{i}\nonumber \\
sc^{a}&=& \frac{g}{2}f^{abc}c^{b} c^{c}\,\, sc=0
\end{eqnarray}

The indices $i,j$ are in the fundamental representation of $SU(2)$, taking values $1$ and $2$. The indices $a$ are in the adjoint representation, taking values $1,2,3$. The non-abelian ghost is $c^{a}$ and the abelian ghost is $c$.

The covariant derivatives and their transformation properties are:

\begin{eqnarray}
(\nabla_{\mu} \phi)_{i}&=& \partial \phi_{i}-ig(T^{a})_{ij}A^{a}_{\mu}\phi_{j} -iea_{\mu}\phi_{i}\nonumber \\
s(\nabla_{\mu} \phi)_{i}&=& -ig (T^{a})_{ij}c^{a} (\nabla_{\mu} \phi)_{j} - iec  (\nabla_{\mu} \phi)_{i}\nonumber \\
(\nabla_{\mu} \phi)^{\dagger}_{i}&=&\partial \phi^{\dagger}_{i}+ ig \phi^{\dagger}_{j}A^{a}_{\mu}(T^{a})_{ji}+iea_{\mu}\phi^{\dagger}_{i}\nonumber \\
s(\nabla_{\mu} \phi)^{\dagger}_{i}&=& ig(\nabla_{\mu} \phi)^{\dagger}_{j}c^{a}(T^{a})_{ji}+iec(\nabla_{\mu} \phi)^{\dagger}_{i}\nonumber \\
sA^{a}_{\mu} &=& -(\partial_{\mu}c^{a} + gf^{abc}A^{b}_{\mu}c^{c}) = - (D_{\mu}c)^{a}\,\,\, sa_{\mu} =-\partial_{\mu}c
\end{eqnarray}

Suppose the action contains a potential term for the scalar fields that induces the following symmetry breaking:

\begin{eqnarray}
\phi_{i}&\rightarrow& \phi_{i} + \nu u_{i},\,\, \nu= \frac{m}{\sqrt{\lambda}}\nonumber \\
\phi^{\dagger}_{i}&\rightarrow& \phi^{\dagger}_{i} + \nu u_{i},\,\, u_{1}=0, u_{2}=1\nonumber \\
v_{1}&=&1, v_{2}=0. u_{i}u_{i}=1, v_{i}v_{i}=1, v_{i}u_{i}=0.
\label{subsphi}
\end{eqnarray}

A typical example is the term:

\begin{equation}
-m^{2}\phi^{\dagger}_{i}\phi_{i} + \frac{\lambda}{2}(\phi^{\dagger}_{i}\phi_{i} )^{2}
\end{equation}

which is invariant under the symmetry defined above.

Thus, a Euclidean gauge action containing all these fields and invariant under the given symmetry is:

\begin{eqnarray}
S&=& \int d^{4}x_{E}\{ \frac{1}{4}F^{a}_{\mu\nu} F^{a\mu\nu} + \frac{1}{4}f_{\mu\nu}f^{\mu\nu} - (\nabla_{\mu} \phi)_{i}(\nabla_{\mu} \phi)^{\dagger}_{i} \nonumber \\
&-&m^{2}\phi^{\dagger}_{i}\phi_{i}  + \frac{\lambda}{2}(\phi^{\dagger}_{i}\phi_{i} )^{2} \}
\end{eqnarray}

where $f_{\mu\nu}$ is the curvature for the abelian field.

As is standard, we substitute the scalar fields as defined in (\ref{subsphi}) by the field plus its vacuum expectation value. This changes the potential and gives mass to the gauge fields. Linear terms appear in the potential, new bilinear terms arise, cubic terms appear, etc. This is standard procedure.

Substituting the scalar fields defined in (\ref{subsphi}) by the field plus its vacuum expectation value, we obtain:

\begin{eqnarray}
s\phi_{i}&=& -ig (T^{a})_{ij}c^{a}\phi_{j}-ig\nu (T^{a})_{ij}c^{a} u_{j}  - iec\phi_{i} -ie \nu c u_{i}\nonumber \\
s\phi^{\dagger}_{i}&=& ig \phi^{\dagger}_{j}c^{a}(T^{a})_{ji} + ig \nu u_{j}c^{a}(T^{a})_{ji} +iec\phi^{\dagger}_{i} +ie\nu c u_{i}\nonumber \\
\nu&=& \frac{m}{\sqrt{\lambda}}
\end{eqnarray}

One easily sees that the new action, obtained after the scalar field redefinition, is invariant under the modified symmetry. The symmetry itself is not broken; rather, the vacuum is. The action defined in the trivial vacuum is invariant under the symmetry in that vacuum, while the action in the non-trivial vacuum is invariant under a transformed expression of the symmetry. However, the two are not equivalent. We now demonstrate that the associated cohomology changes \cite{Amaral:2022cje}\cite{Amaral:2023yti}.

We use a theorem from Ref. \cite{Piguet:1995er} stating that the cohomology of an operator is a subspace of the cohomology of its filtered part. This theorem is the basis of the cohomological reconstruction mechanism \cite{Henneaux:1994lbw}\cite{Barnich:2025jna}\cite{Henneaux:1997mf}. The filtration is defined via number operators that commute with the BRST operator. In this case, we use the simplest filtration, in which the filtered operator contains only one ghost field \cite{Piguet:1995er}.

Before symmetry breaking (trivial vacuum), the filtered operator vanishes for both scalars. After breaking, the filtered operator takes the form:

\begin{eqnarray}
s_{0}\phi_{i}&=& -ig\nu (T^{a})_{ij}c^{a} u_{j}  -ie \nu c u_{i}\nonumber \\
s_{0}\phi^{\dagger}_{i}&=& ig \nu u_{j}c^{a}(T^{a})_{ji} +ie\nu c u_{i}
\end{eqnarray}

Recalling that $T^{a}=\frac{1}{2}\sigma^{a}$ for $a=1,2,3$ and writing $-ie\nu c u_{i} = -ie \nu c \delta_{ij}u_{j}$, we obtain componentwise:

\begin{eqnarray}
s_{0} \phi_{1} &=& -ig\frac{\nu}{2} c^{-}\nonumber \\
s_{0}\phi_{2}&=& ig\frac{\nu}{2} c^{3} -ie\nu c = i\nu c_{z}\nonumber \\
s_{0}\phi^{\dagger}_{1} &=&ig\frac{\nu}{2} c^{+}\nonumber \\
s_{0}\phi^{\dagger}_{2} &=&- ig\frac{\nu}{2} c^{3} +ie\nu c =- i\nu c_{z}\nonumber \\
c^{+} &=& \frac{1}{2}(c^{1}+ic^{2}), \,\,\, c^{-} = \frac{1}{2}(c^{1}-ic^{2})
\end{eqnarray}

We then see that the following combinations are invariant under the filtered operator:

\begin{eqnarray}
s_{0}( \partial_{\mu}\phi_{2} +  \partial_{\mu}\phi^{\dagger}_{2} ) &=&0\nonumber \\
s_{0}( \partial_{\mu}\phi_{2} - \partial_{\mu}\phi^{\dagger}_{2} + 2i\nu Z^{0}_{\mu})&=&0 \nonumber \\
s_{0}(\partial_{\mu}\phi^{\dagger}_{1} + ig\frac{\nu}{2}W^{+}_{\mu})&=&0\nonumber \\
s_{0}(\partial_{\mu}\phi_{1} + ig\frac{\nu}{2}W^{-}_{\mu})&=&0\nonumber \\
W^{+}_{\mu}&=&\frac{1}{2}(A^{1}_{\mu}+iA^{2}_{\mu})\nonumber \\
W^{-}_{\mu}&=&\frac{1}{2}(A^{1}_{\mu}-iA^{2}_{\mu})\nonumber \\
Z^{0}_{\mu}&=& \frac{g}{2}A^{3}_{\mu} - e a_{\mu}\nonumber \\
\gamma_{\mu}&=& \frac{g}{2}A^{3}_{\mu} + e a_{\mu}, 
\end{eqnarray}

This relation can be lifted to the full operator $s$, completing the cohomology. The fundamental point is that this combination behaves essentially as matter, not as a gauge field. The question remains how to introduce an appropriate gauge fixing that eliminates, at least on-shell, the Gribov problem.

\section{Practical notation and the Morse functional}

From this cohomological construction, it becomes clear that the entire analysis follows directly from the sector:

\begin{eqnarray}
s_{0}\phi_{i}&=& -i\nu (g (T^{a})_{ij}c^{a}  + e  c \delta_{ij})u_{j}\nonumber \\
s_{0}\phi^{\dagger}_{i}&=& i \nu u_{j}(gc^{a}(T^{a})_{ji} +e c \delta_{ji})
\end{eqnarray}

This allows us to define the new variable $c_{ij}= (gc^{a}(T^{a})_{ji} +e c \delta_{ji})$, giving:

\begin{eqnarray}
s_{0}\phi_{i}&=& -i\nu c_{ij}u_{j}\nonumber \\
s_{0}\phi^{\dagger}_{i}&=& i \nu u_{j}c_{ji}
\end{eqnarray}

One finds that the filtrations correspond to:

\begin{eqnarray}
u_{i}c_{ij}u_{j}&=& -\frac{g}{2}c^{3}+ec= -c_{Z}\nonumber \\
v_{i}c_{ij}u_{j}&=& gc^{-}\nonumber \\
u_{i}c_{ij}v_{j}&=& gc^{+}\nonumber \\
v_{i}c_{ij}v_{j}&=& \frac{g}{2}c^{3}+ec = c_{\gamma}\nonumber \\
\end{eqnarray}

The same reasoning applies to the gauge sector $(A^{\mu})_{ij}= (gA^{\mu a}(T^{a})_{ji} +e a^{\mu} \delta_{ji})$:

\begin{eqnarray}
u_{i}(A^{\mu})_{ij}u_{j}&=& -Z^{\mu}\nonumber \\
v_{i}(A^{\mu})_{ij}u_{j}&=& gW^{\mu -}\nonumber \\
u_{i}(A^{\mu})_{ij}v_{j}&=& gW^{\mu +}\nonumber \\
v_{i}(A^{\mu})_{ij}v_{j}&=& \gamma^{\mu}\nonumber \\
\end{eqnarray}

For the original BRST symmetries (without a scalar vacuum expectation value), we have:

\begin{eqnarray}
sc_{ij}&=& -i c_{il}c_{lj}\nonumber \\
s (A_{\mu})_{ij}&=& - (\partial_{\mu}c_{ij} + i c_{il}(A_{\mu})_{lj} -i (A_{\mu})_{il}c_{lj})\nonumber \\
s\varphi_{i}&=& -ic_{il}\varphi_{l} -i \nu c_{il}u_{l} \nonumber \\
s\varphi^{\dagger}_{i}&=& i \varphi^{\dagger}_{l}c_{li} + i \nu u_{l} c_{li}\nonumber \\
\end{eqnarray}

For an appropriate gauge choice, we have:

\begin{equation}
s\overline{c}_{ij}=i b_{ij},\,\,\,\, s  b_{ij}=0
\end{equation}

The reasoning for $SU(2)\times U(1)$ extends trivially to $SU(N)\times U(1)$. The fundamental representation has $N$ components, with projection vectors $u_{i}$ and $v_{i}$ defined in (\ref{subsphi}) having $N$ components. A symmetry breaking that projects onto a single component of the vector leaves an invariant $SU(N-1)$ subgroup. A second symmetry breaking can then give mass to the fields in the subgroup. This reasoning is important for the Morse problem discussed below.

We now construct a generating functional to study the Morse problem. The Morse problem is an extremum problem for functionals and is directly linked to the Gribov problem. Specifically, the Gribov problem concerns the eigenvalues of the operator arising from the gauge fixing condition.

We now present the antiBRST transformation:

\begin{eqnarray}
\overline{s}\overline{c}_{ij}&=&-i\overline{c}_{il}\overline{c}_{lj}\nonumber \\
\overline{s}b_{ij}&=& ib_{il}\overline{c}_{lj}- i \overline{c}_{il}b_{lj}\nonumber \\
\overline{s}c_{ij}&=& -i(c_{il}\overline{c}_{lj} + \overline{c}_{il}c_{lj} + b_{ij})\nonumber \\
\overline{s} (A_{\mu})_{ij}&=& - (\partial_{\mu}\overline{c}_{ij} + i \overline{c}_{il}(A_{\mu})_{lj} -i (A_{\mu})_{il}\overline{c}_{lj})\nonumber \\
\overline{s}\varphi_{i}&=& -i\overline{c}_{il}\varphi_{l} -i \nu \overline{c}_{il}u_{l} \nonumber \\
\overline{s}\varphi^{\dagger}_{i}&=& i \varphi^{\dagger}_{l}\overline{c}_{li} + i \nu u_{l} \overline{c}_{li}\nonumber \\
\{s,\overline{s}\}&=&0
\end{eqnarray}

To illustrate this process, we write a functional of ultraviolet dimension $2$, ghost number $0$, Lorentz and group invariant, containing only the gauge field. This functional generates the Landau gauge.

Define the functional $\mathcal{I}$ as:

\begin{equation}
\mathcal{I}= \int d^{4}x \frac{1}{2}  (A_{\mu})_{ij} (A^{\mu})_{ji}
\end{equation}

We then find:

\begin{eqnarray}
S_{gf}&=& s\overline{s} \mathcal{I} = \int d^{4}x (ib_{ij}\partial^{\mu} (A_{\mu})_{ji} + \overline{c}_{ji}\partial^{\mu} (\partial_{\mu}c_{ij} + i c_{il}(A_{\mu})_{lj} -i (A_{\mu})_{il}c_{lj}) )\nonumber \\
\frac{\delta S_{gf}}{\delta b_{ij}}&=& i \partial^{\mu} (A_{\mu})_{ji}\nonumber \\
\frac{\delta S_{gf}}{\delta \overline{c}_{ij}}&=&\partial^{\mu} (\partial_{\mu}c_{ji} + i c_{jl}(A_{\mu})_{li} -i (A_{\mu})_{jl}c_{li}) )
\label{Isimples}
\end{eqnarray}

Substituting the gauge condition $\frac{\delta S_{gf}}{\delta b_{ij}}=0$ into the antighost equation yields the on-shell Gribov operator:

\begin{equation}
\partial^{\mu} \partial_{\mu}c_{ji} + i \partial^{\mu} c_{jl}(A_{\mu})_{li} - i  (A_{\mu})_{jl}\partial^{\mu}c_{li}=0.
\end{equation}

Given a scalar functional of ultraviolet dimension $2$ containing only the gauge field, we obtain the gauge condition as a simple Morse problem. One may also introduce the product $\overline{c}_{ij}c_{ji}$ without changing the on-shell Gribov condition. In an action containing scalar fields, one could also include them. Thus the most general functional of ultraviolet dimension $2$ and ghost number zero is:

\begin{equation}
\mathcal{I}= \int d^{4}x  \left(\frac{1}{2}  (A_{\mu})_{ij} (A^{\mu})_{ji} + a \overline{c}_{ij}c_{ji} + \beta \varphi^{\dagger}_{i}\varphi_{i}\right).
\end{equation}

Note that the last term has vanishing BRST and anti-BRST variations in the trivial vacuum, but not in the broken vacuum obtained from (\ref{Isimples}):

\begin{eqnarray}
s\overline{s}\int d^{4}x  (\beta \varphi^{\dagger}_{i}\varphi_{i})&=& \int d^{4}x  (-\beta\nu u_{l}b_{li}\varphi_{i} + \beta\nu \varphi^{\dagger}_{i}b_{il}u_{l} \nonumber \\
&-& \beta \nu \overline{c}_{li}c_{il}\varphi_{l} + \beta\nu \varphi^{\dagger}_{l}c_{li}\overline{c}_{il}u_{l}\nonumber \\
&-&\beta \nu^{2}u_{l}\overline{c}_{li}c_{il}u_{l} + \beta \nu^{2} u_{l}c_{li}\overline{c}_{il}u_{l}).
\end{eqnarray}

Without presenting all intermediate steps (the derivation of the gauge fixing action and the Lagrange multiplier equations), we obtain:

\begin{equation}
\partial^{\mu} \partial_{\mu}c_{ji} + i \partial^{\mu} c_{jl}(A_{\mu})_{li} - i  (A_{\mu})_{jl}\partial^{\mu}c_{li} - \nu^{2}\beta(u_{i}c_{jl}u_{l}  + u_{l}c_{li}u_{j})=0.
\end{equation}

The last term is a mass term. Therefore, in the broken phase (i.e., for energies compatible with the broken vacuum), we find that the on-shell Gribov problem does not exist. The Gribov sector (identical to the massless case) has eigenvalues smaller than the term $\nu^{2}\beta(u_{i}c_{jl}u_{l}  + u_{l}c_{li}u_{j})$ by definition of symmetry breaking. One might imagine that topological effects such as instantons could alter this result. However, from a perturbative point of view, the on-shell problem is solved.

This result is independent of the values of $a$ and $\beta$. Different choices of $a$ and $\beta$ lead to different gauge fixings. Thus, any gauge fixing built from a Morse functional in a theory with spontaneous symmetry breaking (where renormalizability and unitarity are imposed) solves the Gribov problem. This includes 't Hooft-type gauges.

\section{Conclusion}

We have shown that spontaneous symmetry breaking modifies the BRST cohomology of a gauge theory, creating matter-like combinations of gauge and scalar fields. Using a Morse functional of dimension $2$ and ghost number zero, we constructed a gauge fixing that yields an on-shell Gribov operator containing a positive mass term proportional to $\nu^2\beta$. This mass term lifts the zero modes responsible for Gribov copies, eliminating the on-shell Gribov problem in the broken phase for perturbative energies below the symmetry breaking scale.

The construction applies to $SU(2)\times U(1)$ and extends to $SU(N)$. A complete spectral proof of positivity is left for future work. The structure presented may indicate a deeper relation between renormalizability, unitarity and spontaneous symmetry breaking in non-abelian gauge theories.

\end{document}